\documentclass[a4paper,aps,twocolumn]{revtex4-2}
\usepackage{amsmath,amssymb,setspace,graphicx,xcolor,textcomp}
\usepackage{epstopdf}
\usepackage[T1]{fontenc}
\usepackage{hyperref}

\usepackage{url}

\usepackage{graphicx}
\usepackage{bm}
\usepackage{amsmath}
\usepackage{xcolor}
\usepackage{hyperref}
\usepackage{url}

\graphicspath{{./}}

\begin{document}

\title{Acceleration-based clustering reveals frequent gait switching in sprint sled dogs}

\author{Benjamin Seleb}
\thanks{This article has been accepted for publication in \textit{Integrative and Comparative Biology}, Published by Oxford University Press.}
\affiliation{Interdisciplinary Graduate Program in Quantitative Biosciences, Georgia Institute of Technology, Atlanta, Georgia, United States}

\author{Saad Bhamla}
\email{saad.bhamla@colorado.edu}
\affiliation{School of Chemical and Biomolecular Engineering, Georgia Institute of Technology, Atlanta, Georgia, United States}
\affiliation{BioFrontiers Institute and Department of Chemical and Biological Engineering, University of Colorado Boulder, Boulder, Colorado, United States}

\date{\today}

\begin{abstract}
Continuous video is difficult to obtain during field studies of sprint sled dogs, limiting analysis of stride-to-stride variation during load-pulling gallop. We developed an acceleration-based pipeline to identify recurrent stride states from harness-mounted tri-axial accelerometers without manual gait labels. Using multivariate dynamic time warping, manifold embedding, and density-based clustering, we analyzed more than 20,000 strides from a 10-dog team and identified recurrent, dog-specific stride states. In one previously annotated individual, acceleration-derived states were broadly consistent with manually labeled gallop patterns. Across dogs, transitions between stride states were frequent, with substantial inter-individual variation and limited evidence of strong team-level coordination. A simple logistic model based on local tugline-force timing and magnitude had weak predictive power for transition events. These results suggest that sprint sled dog gallop occupies a variable set of nearby stride states and that local tugline-force fluctuations alone do not explain the observed switching.
\end{abstract}


\maketitle

\section{Introduction}\label{sec:labeled}

High-speed gallop is often treated as a stereotyped gait whose expression is governed mainly by speed. Sprint sled dogs provide a useful system for testing that view because they gallop under sustained external load, with mechanical coupling to the vehicle transmitted through the tugline. Sprint sled dog racing involves teams of dogs maintaining sustained speeds of roughly 25--40 km h$^{-1}$ over distances of tens of kilometers \cite{thorsrud2021description}. At these speeds, sled dogs predominantly use a gallop \cite{ding2025gait}, in contrast to longer-distance sled dog racing where variable speed ranges enable a greater diversity of gaits. 

In previous work \cite{ding2025gait}, we used high-speed video and manually annotated limb stance timing to identify multiple galloping patterns (both rotary and transverse gallops) in sprint sled dogs. Surprisingly, dogs frequently switched between these galloping patterns within a few strides despite only small changes in speed and stride duration. This observation contrasts with the traditional view that gait transitions are primarily regulated by speed, with animals shifting gaits to minimize energetic cost \cite{Hoyt1981gaitenergetic}.

However, gait transitions may also arise from mechanical perturbations. For example, horses have been shown to transition from trot to gallop when peak musculoskeletal forces reach a critical threshold, rather than when energetically optimal \cite{Farley1991trigger}. Similarly, mechanical perturbations can elicit shifts from trot to bound in rodents \cite{vahedipour2018uncovering}. In dogs, it has also been suggested that sufficiently large disturbances may induce transitions between galloping modes \cite{wilshin2020dog}. Together, these findings suggest that external mechanical influences, such as those associated with load pulling, could trigger the observed switching between galloping modes in sled dogs.

Ideally, this hypothesis would be tested using long, synchronized records of gait and tugline force. However, available video datasets are short and contain too few transitions to examine potential drivers of switching such as tugline dynamics. Obtaining continuous side-view video in field settings is impractical for several reasons. Sprint sled dog teams travel on narrow trails at race speeds, limiting stationary cameras to capture windows of only a few seconds per pass. While aerial solutions offer a promising approach for collecting behavioral data from fast-moving or widely dispersed animals \cite{Koger_2023}, drone footage may not achieve the spatial resolution needed to resolve individual limb contacts at these speeds and distances. Within-team occlusion further compounds these challenges, as dogs frequently obscure one another from any fixed vantage point. Animal-borne sensors (biologgers)---typically carrying GPS, inertial measurement units, or other sensors---offer an alternative strategy to track movement and physiology throughout the run \cite{Hayati_2019, Dewhirst_2017, Fehlmann_2017}. Biologgers have been used to study locomotor behavior and social dynamics across a wide range of taxa in field conditions where direct observation or video would otherwise be infeasible \cite{Brown_2013,Tuia_2022,King_2018}. We therefore asked whether stride structure could instead be recovered directly from biologger acceleration signals.


Here we use harness-mounted tri-axial acceleration to identify recurrent stride states without manual gait labels. We first assess the approach in a previously annotated individual and then apply it to long recordings from a 10-dog sprint team. This allows us to quantify stride-state switching over thousands of strides and to test whether local tugline-force features are associated with transition events.


\section{Methods}

\subsection{Gait patterns form distinct acceleration trajectories}

\begin{figure}[h]
\centering
\includegraphics[width=0.5\textwidth]{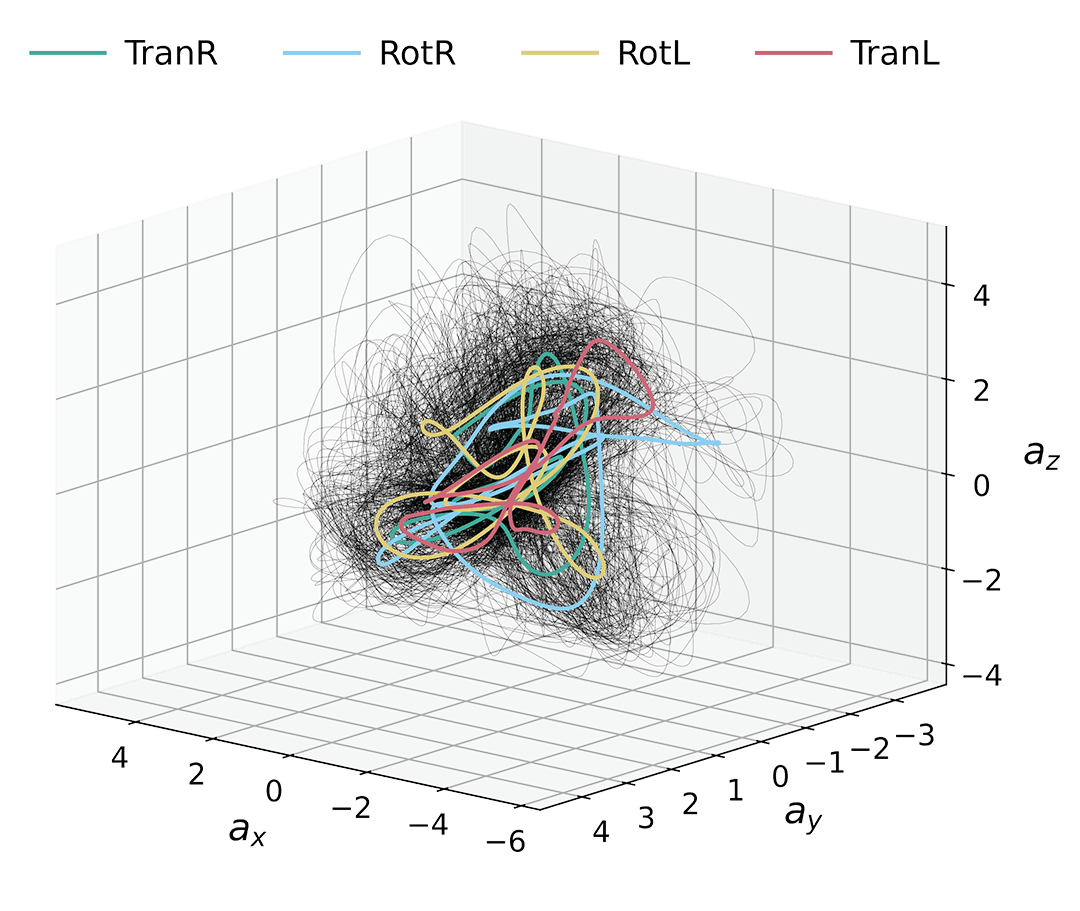}
\caption{\label{fig:labeled_limit_cycle} \textbf{Gait patterns correspond to unique amplitude modes in acceleration space.} Acceleration trajectories from 190 manually labeled strides ($\sim$70\,s) in Individual 1, plotted in acceleration space (axes in $g$). Gray traces show individual spline-interpolated, phase-normalized strides; colored curve shows the phase-averaged trajectory for each manually labeled gallop pattern: \textit{TranL} (transverse, left-leading), \textit{TranR} (transverse, right-leading), \textit{RotL} (rotary, left-leading), and \textit{RotR} (rotary, right-leading). Mean trajectories reveal slightly offset limit cycles in acceleration space.
}
\end{figure}

We first asked whether manually labeled gallop patterns occupy distinguishable regions of acceleration space. Using 190 consecutive strides from a previously analyzed, video-labeled dog (``Individual 1'')~\cite{ding2025gait}, we plotted each stride as a trajectory in three-dimensional acceleration space (Fig.~\ref{fig:labeled_limit_cycle}). Strides were segmented between successive hind-limb touchdown events to match the conventions of the labeled dataset. Grouping strides by their classified gait pattern reveals that repeated strides trace recurrent trajectories, forming distinct limit cycles in acceleration space. To highlight these differences, trajectories belonging to each gait pattern were time-aligned and averaged to obtain mean stride cycles. The resulting mean trajectories are shown as colored curves in Fig.~\ref{fig:labeled_limit_cycle}. Rotary and transverse gallops occupy overlapping but slightly displaced regions of acceleration space, indicating that acceleration waveform geometry carries information about labeled gait pattern. This separation motivates the use of unsupervised methods to identify recurring stride states directly from acceleration dynamics \cite{Brown_2018,couzin2023emerging}.

\subsection{Data collection and preprocessing}

Data were collected from a 10-dog sprint team during a continuous training run, with each dog contributing more than 2,000 consecutive strides. The team pulled a motorized ATV carrying two passengers.
Each dog was fitted with a custom harness-mounted biologger positioned near the withers on a standard X-back harness. The logger recorded tri-axial acceleration at 125 Hz and tugline force at 62 Hz throughout the run. Additional details on sensor design, firmware, and hardware are provided in the Data and Materials Availability section. A GPS unit mounted on the ATV recorded team position throughout the run, from which instantaneous speed and heading were derived. Periods when GPS speed fell below 3\,m\,s$^{-1}$ were excluded prior to analysis, accounting for 1\% of the total run. During the remaining interval, team speed was maintained at race-like speeds ($5.97 \pm 0.65$\,m\,s$^{-1}$; range 3.01--7.15\,m\,s$^{-1}$). This involved predominantly straight-line running, with a mean heading angular rate of 1.43\,deg\,s$^{-1}$ and only 5.8\% of the analyzed interval exceeding 5\,deg\,s$^{-1}$.

Raw data were screened for completeness and then processed as follows.

\noindent
\textbf{Acceleration.} Tri-axial acceleration was high-pass filtered using a fourth-order zero-phase Butterworth filter (cutoff 0.5 Hz) to remove quasi-static/gravity components \cite{Dewhirst_2017}. The dorsoventral component ($a_z$) was z-normalized prior to stride segmentation and phase computation.

\noindent
\textbf{Force.} For identifying force peak arrival times, force was lightly smoothed with a Savitzky--Golay filter (\texttt{scipy.signal.savgol\_filter}, window 11, poly 3) to suppress spurious local peaks, then resampled by linear interpolation to 125 Hz for alignment with acceleration.

\noindent
\textbf{Time.} All data streams were synchronized to GPS time, then trimmed to the common overlap across dogs.

\subsection{Stride segmentation}

Stride boundaries were identified using a marker-event segmentation approach based on zero-crossings in the dorsoventral acceleration signal ($a_z$). Let these crossing times be denoted $t_k$, where $k$ indexes successive strides. Each isolated stride was then represented as a three-channel subsequence (Fig.~\ref{fig:DTW1}A).

Steep positive zero crossings---instances where $a_z$ changed sign from negative to positive, identified algorithmically---were taken as stride onset times $t_k$. To prevent false detections caused by small oscillations near zero, only crossings separated by at least 0.25 s (corresponding to a maximum frequency of 4 Hz) were retained. Segments shorter than two sampling periods were discarded automatically.

\begin{figure}[h]
\centering
\includegraphics[width=0.45\textwidth]{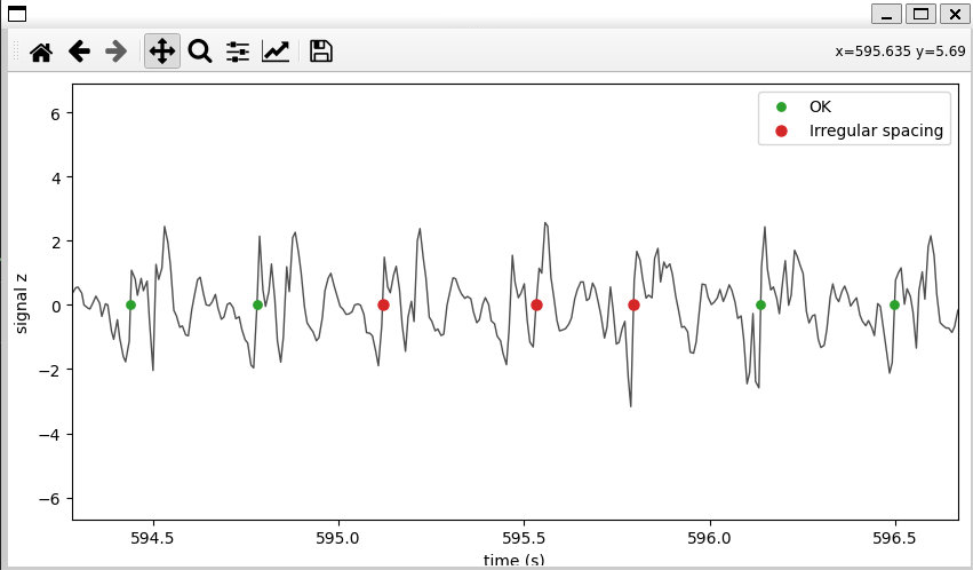}
\caption{\label{fig:segmentationUI} \textbf{Segmentation validation interface.} Interface used to validate and correct algorithmically identified stride markers. Irregularly spaced markers are highlighted for manual inspection.}
\end{figure}

All automatically detected crossings were visually inspected in an interactive Python interface that displayed the $a_z$ trace with overlaid crossing markers. False detections were corrected manually using a click-and-drag editor that allowed crossing markers to be repositioned, added, or deleted (Fig.~\ref{fig:segmentationUI}).

To compare subsequences belonging to different strides, we define a linearly increasing phase that advances by $2\pi$ each stride,

\begin{equation}
    \label{eq:phase}
   \phi(t) = 2\pi k + 2\pi\frac{t - t_k}{t_{k+1} - t_k}.
\end{equation}

This representation is equivalent to normalizing time by stride duration, while preserving the periodic structure of the stride dynamics.

\subsection{Stride-space representation}

To compare these stride subsequences while allowing for small temporal distortions, we computed pairwise distances between waveforms using multivariate dynamic time warping (MDTW) \cite{shokoohi2017generalizing, mearns2020deconstructing}, implemented with the Python package \texttt{dtaidistance}. Distances were computed separately for each dog.

For two stride subsequences $\mathbf{A}_i$ and $\mathbf{A}_j$ of length $T_i$ and $T_j$, MDTW finds the minimal-cost alignment path $\epsilon = \{(t_i, t_j)\}$ that minimizes
\begin{equation}
\mathbf{D}_{ij} = \min_{\epsilon} \sum_{(t_i,t_j)\in \epsilon}
\lVert \mathbf{a}_i(t_i) - \mathbf{a}_j(t_j)\rVert_2.
\label{eq:MDTW}
\end{equation}

\begin{figure*}[ht!]
\centering
\includegraphics[width=1\textwidth]{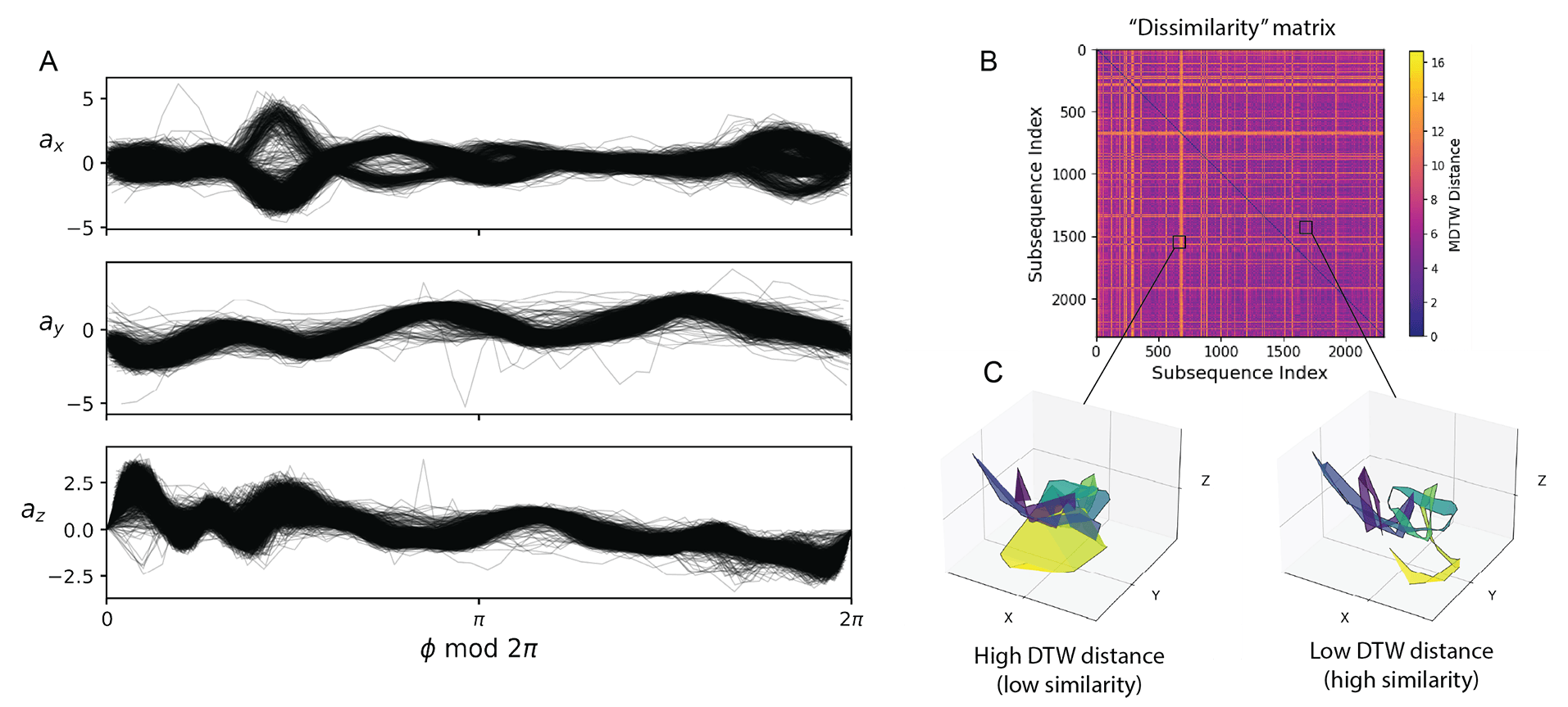}
\caption{\label{fig:DTW1} \textbf{Computation of MDTW distances for a representative dog from the 10-dog dataset.}
(A) Segmented tri-axial acceleration  subsequences for successive strides, aligned by stride phase $\phi(t_k)$ (Eq.~\ref{eq:phase}).
(B) Pairwise MDTW distances among $\sim$2000 strides ($\sim$700\,s) from a single dog produces a distance matrix.
(C) Example 3-axis alignments of stride pairs: thicker ribbons represent larger distances (dissimilar dynamics) while thin ribbons correspond to smaller MDTW distances (similar strides); ribbon color indicates time/phase progression through the stride (early to late).}
\end{figure*}

The resulting matrix $\mathbf{D}$ encodes the pairwise dissimilarities between all stride pairs within a dog (Fig.~\ref{fig:DTW1}B). Smaller $\mathbf{D}_{ij}$ values indicate more dynamically similar stride trajectories (Fig.~\ref{fig:DTW1}C).

The dissimilarity matrix $\mathbf{D}$ was embedded into a two-dimensional manifold using Uniform Manifold Approximation and Projection (UMAP) \cite{mcinnes2020uniform}, implemented with the Python package \texttt{umap-learn} (parameters: $n_{\mathrm{neighbors}}=12$, $\mathrm{min\_dist}=0.22$). In this representation, dynamically similar strides lie close together in the embedded space, forming clusters that correspond to recurring stride dynamics.

To identify these recurring stride states, we applied density-based clustering using DBSCAN \cite{ester1996density}, implemented with \texttt{sklearn.cluster} (parameters: $\epsilon = 0.6$ and $\mathrm{min\_samples}=10$). Each stride $k$ was assigned a cluster label $s_k \in \{1,2,\dots,S\}$. Sensitivity of the identified stride states and switching statistics to these parameter choices is assessed in Appendix~\ref{app:robustness}.

\subsubsection{Stride syllables and switching dynamics}

The cluster assignment sequence $s_k$ can be interpreted as a series of discrete behavioral syllables \cite{wiltschko2015mapping}, providing a compact representation of real-time stride-state variation.

Because each dog's embedding is stochastic and unique, stride clusters are not directly comparable across individuals (e.g., state~0 for one dog need not correspond to state~0 for another). To enable cross-dog comparisons, we reduced each syllable sequence to a binary switching series indicating when a transition between stride states occurred.

If $s_k$ denotes the cluster assignment of stride $k$, a binary switch indicator was defined as
\begin{equation}
b_k =
\begin{cases}
1, & \text{if } s_{k+1} \neq s_k,\\[4pt]
0, & \text{otherwise,}
\end{cases}
\label{eq:switch_indicator}
\end{equation}

where $b_k=1$ marks a transitional stride and $b_k=0$ indicates stride-to-stride stability. This conversion produces a binary time series $\mathbf{b} = [b_1, b_2, \dots, b_K]$ analogous to a spike train in neuroscience, where the ``spikes'' here represent stride-state transitions.

\section{Results}

\subsection{Unsupervised clustering reveals stride states}

\begin{figure*}[ht!]
\centering
\includegraphics[width=1\textwidth]{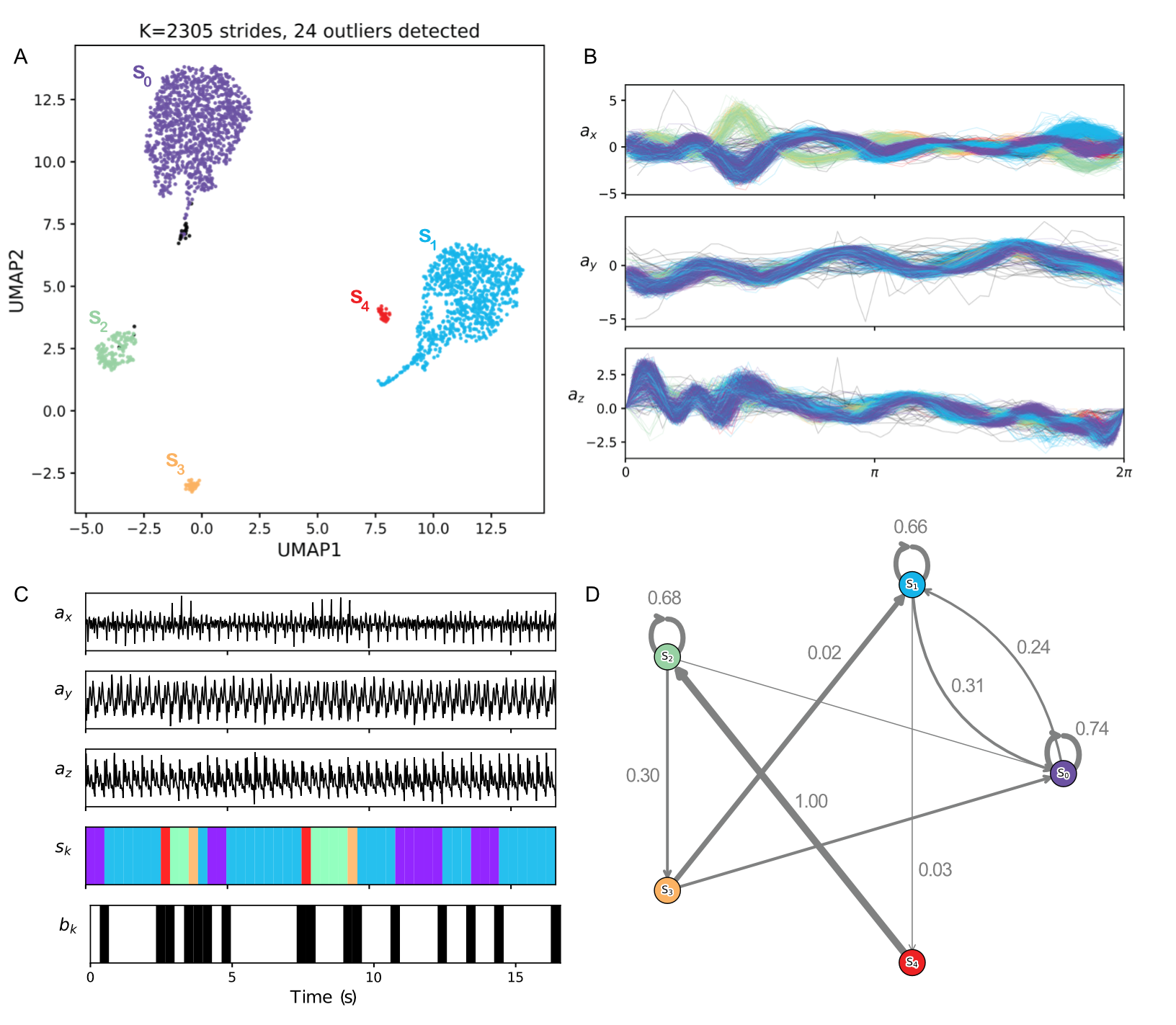}
\caption{\label{fig:DTW2} \textbf{Stride-space representation for a single dog.} (A) Two-dimensional UMAP embedding of the per-stride MDTW distance matrix.  Each point represents a stride, colored by cluster assignment $s$; black points denote outliers (top 1\% of distances).
(B) The same strides mapped back on to acceleration traces and  recolored by  cluster label, revealing unique, overlapping amplitude modes.
(C) Temporal mapping of stride clusters: the stride-state sequence $s_k$ (upper panel) and binary switch indicator $b_k$ (lower panel) for a representative segment. Stride-state assignments reduce the acceleration signal to discrete behavioral syllables, and transitions between syllables define the switch train.
(D) Transition network for the same individual as panels (A)--(C); edge widths and labels denote transition probabilities and self-loops indicate persistence within a state.}
\end{figure*}

Applying the pipeline described in Methods across the ten dogs, this procedure identified between four and seven recurrent stride states per individual (Table~\ref{tab:dwell_stats}). Mapping cluster labels back onto the phase-aligned acceleration traces reveals distinct braid-like trajectories representing each dog's repertoire of stride modes (Fig.~\ref{fig:DTW2}B).

\begin{table*}[!t]
\centering
\caption{Per-dog stride and switching statistics.}
\begin{tabular}{cccccccc}
\hline
ID & Strides & Switches & States & Mean dwell (strides) & Mean dwell (s) & $C_V$ \\
\hline
1 & 2305 & 741 & 6 & 3.11 & 1.03 & 1.00 \\
2 & 2268 & 775 & 5 & 2.92 & 0.99 & 1.35 \\
3 & 2282 & 394 & 5 & 5.78 & 1.94 & 1.66 \\
4 & 2279 & 605 & 6 & 3.76 & 1.26 & 1.92 \\
5 & 2269 & 1001 & 6 & 2.26 & 0.77 & 1.32 \\
6 & 2082 & 179 & 6 & 11.57 & 4.26 & 1.17 \\
7 & 2282 & 538 & 7 & 4.23 & 1.42 & 2.61 \\
8 & 2425 & 57 & 5 & 41.81 & 13.21 & 2.39 \\
9 & 2178 & 704 & 7 & 3.09 & 1.09 & 1.79 \\
10 & 2124 & 773 & 4 & 2.74 & 0.99 & 1.31 \\
\hline
\end{tabular}
\label{tab:dwell_stats}
\end{table*}

\subsubsection{Validation with labeled data}

To verify that the unsupervised clusters correspond to identifiable gait patterns, we compared cluster assignments $s_k$ with the manually labeled gait identities $g_k$ from Individual~1. For each gait label $g$, the dominant cluster $c^*(g)$ was identified, and cluster purity was computed as

\begin{equation}
P_g = \frac{1}{n_g}\sum_{k:g_k=g}\mathbf{1}[c_k=c^*(g)],
\label{eq:purity}
\end{equation}

where $n_g$ denotes the number of strides assigned label $g$.

The weighted mean purity across labels was $0.96$, indicating that the unsupervised embedding reliably recovered gait-specific structure rather than reflecting other sources of variation or noise.

\subsection{Frequent stride-state switching during sustained running}

\subsubsection{Individual dwell statistics}

For each dog, contiguous non-switch segments in $\mathbf{b}$ define \emph{dwell periods} of length $L_i$, corresponding to the number of strides between consecutive stride-state transitions. This measure is analogous to the interspike interval used in neural data analysis \cite{kreuz2009measuring} and captures how long a stride state persists before changing. We quantified the regularity of switching using the coefficient of variation of dwell periods:
\begin{equation}
\mathrm{C_V}_{\mathrm{dwell}} =
\frac{\mathrm{std}(L_i)}{\mathrm{mean}(L_i)}.
\label{eq:CV}
\end{equation}
A value of $\mathrm{C_V}=1$ corresponds to Poisson-like random switching, $\mathrm{C_V}<1$ indicates regular switching, and $\mathrm{C_V}>1$ indicates bursty or irregular transitions. Across individuals, stride-state switching exhibited substantial variability (Fig.~\ref{fig:raster}C, Table~\ref{tab:dwell_stats}). Mean dwell periods ranged from roughly 2--12 strides for most dogs, with one particularly stable individual maintaining dwell periods exceeding 40 strides between transitions (dog~8; no transitions were observed during the 100\,s segment shown in Fig.~\ref{fig:raster}B). The coefficient of variation ranged from 1.0--2.6, indicating irregular or bursty switching rather than regular or periodic transitions.

\subsubsection{Collective switching dynamics}
\begin{figure*}[ht!]
\centering
\includegraphics[width=1\textwidth]{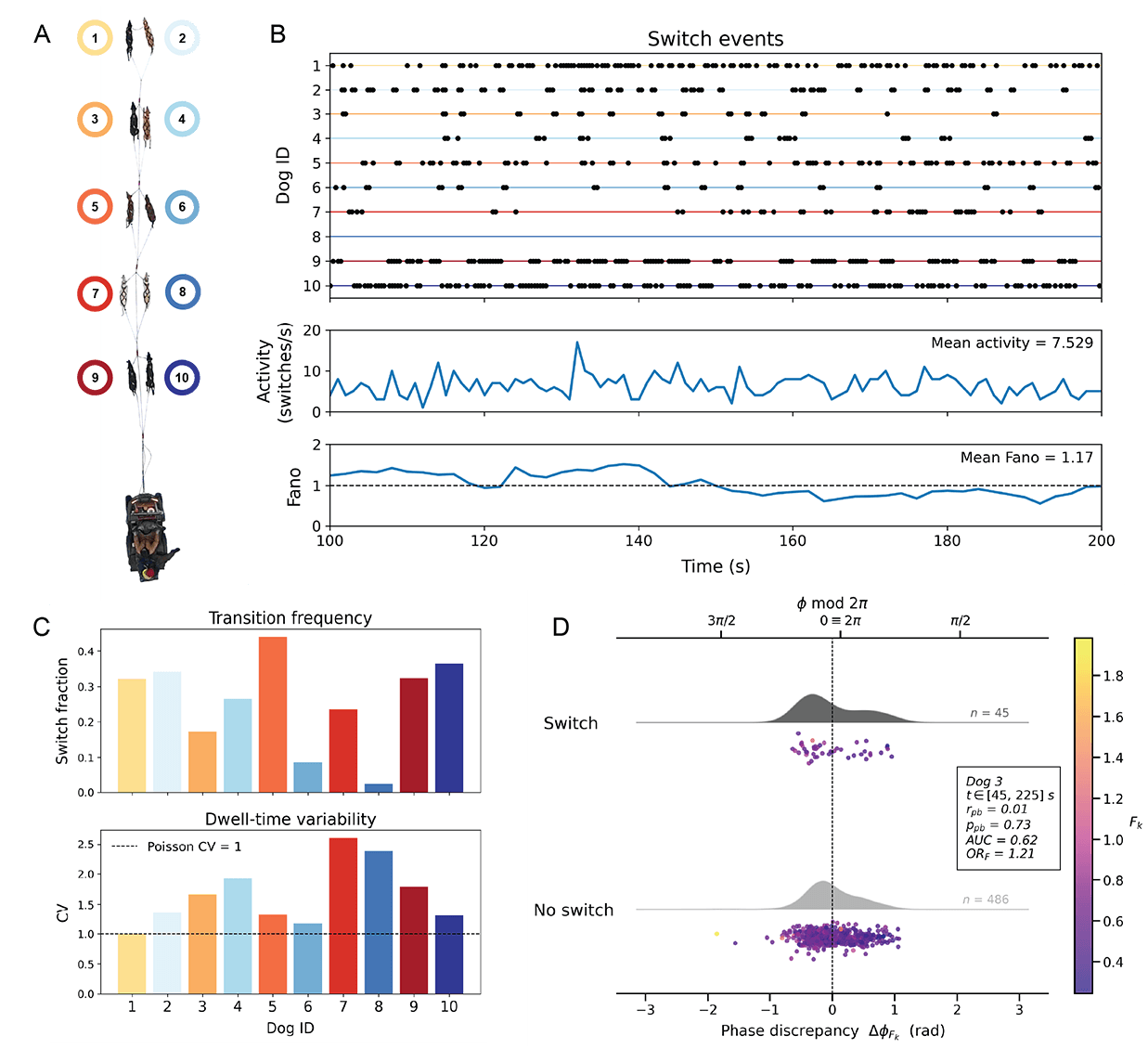}
\caption{\label{fig:raster} \textbf{Teamwide stride-state switching dynamics.}
(A) 10-dog team with color-coded dog identities.
(B) Raster plot of switch events over a 100~s segment, analogous to a neural spike raster: each row corresponds to one dog, and black dots indicate switches between stride states.
Below, the team's total switch count per second and the sliding-window Fano factor reveal that switching is temporally irregular yet generally asynchronous.
(C) Per-dog summary statistics showing transition frequency (fraction of transitional strides) and dwell-time variability (coefficient of variation).
(D) Logistic regression example for one dog, testing whether tugline perturbations can predict switching.  Each point represents a stride colored by peak force~$F_k$; the horizontal axis shows the phase discrepancy, $\Delta \phi_{F_k}$ defined as the deviation of the force peak phase from the mean peak phase across strides, $\Delta\phi_{F_k} = \phi_{F_k} - \bar{\phi}_F$. Values near zero correspond to force peaks occurring at the average timing within the stride, while negative and positive values indicate earlier and later peaks, respectively. Switch (1) or no-switch (0) outcomes are separated by category.}
\end{figure*}

We next examined whether stride-state transitions occurred in a coordinated manner across the 10-dog team (Fig.~\ref{fig:raster}A,B).
Each dog's binary switch train was treated as a point process, and switching events were aggregated across dogs using 1\,s time bins.
The instantaneous population switching activity was defined as

\begin{equation}
\mathcal{R}(t) = \frac{1}{\Delta t} \sum_{n=1}^{N_{dogs}} \mathcal{N}_n(t, t+\Delta t),
\label{eq:pop_activity}
\end{equation}

where $\mathcal{N}_n(t,t+\Delta t)$ denotes the number of switches produced by dog $n$ within the window of width $\Delta t$. Here, $\mathcal{R}(t)$ reflects the absolute number of switch events across the team each second.

To quantify variability in this population activity, we computed the Fano factor \cite{rajdl2020fano},

\begin{equation}
\mathcal{F} =
\frac{\mathrm{Var}[\mathcal{N}(t, t+\Delta t)]}
     {\mathrm{E}[\mathcal{N}(t, t+\Delta t)]}.
\label{eq:fano}
\end{equation}

Under idealized Poisson switching, a Fano factor near $1$ indicates independent switching, and $\mathcal{F}>1$ indicates correlated or partially synchronized transitions ($\mathcal{F}=N_{dogs}=10$ for perfect synchrony). A sliding-window estimate $\mathcal{F}(t)$ was computed using 10\,s windows advanced in 1\,s steps to visualize temporal fluctuations in collective variability.

Across the team, dogs switched stride states frequently ($\bar{\mathcal{R}}=7.54$), with mean dwell periods of only a few strides for most individuals. The mean Fano factor was $\bar{\mathcal{F}}=1.17$, slightly above the Poisson expectation. This suggests at most weak temporal coordination among dogs, with most stride-state transitions occurring independently. We note that inter-individual heterogeneity in switching rates (Table~\ref{tab:dwell_stats}) can produce modest positive deviations from $\mathcal{F} = 1$ even for fully independent processes, so the small observed excess may partly reflect rate heterogeneity rather than genuine coordination.


\subsection{Tugline forces weakly predict switching}

To test whether stride-state switching reflects a response to mechanical forcing, we examined whether the timing and magnitude of tugline tension around stride boundaries predict stride-state transitions. Switch probability was modeled using logistic regression with circular (sine--cosine) encoding of the phase of the peak force $\phi_{F_k}$ and the peak force magnitude $F_k$:

\begin{equation}
\mathrm{logit}\,P(y_k=1)
= \beta_0 + \beta_{\cos} \cos \phi_{F_k} + \beta_{\sin} \sin \phi_{F_k} + \beta_F F_k,
\label{eq:logit}
\end{equation}
where $y_k=1$ indicates a stride-state transition between strides $k$ and $k+1$, and $\beta_0$, $\beta_{\cos}$, $\beta_{\sin}$, $\beta_F$ are fitted regression coefficients.

Force magnitude $F_k$ was z-scored prior to fitting. After filtering edge strides and removing outliers, a total of $n_{\text{events}}$ stride-boundary events were retained for analysis of each individual.
Models were fit by maximum likelihood, as implemented in \texttt{scikit-learn}'s \texttt{LogisticRegression} function \cite{pedregosa2011scikit}.
Model performance was evaluated using the receiver operating characteristic (ROC) curve \cite{fawcett2006introduction}, which quantifies the trade-off between true and false positive rates across different decision thresholds.

The area under the ROC curve ($\mathrm{AUC}$) provides a threshold-independent measure of discriminability, where $0.5$ indicates chance-level performance and $1.0$ indicates perfect prediction. The effect of force magnitude was expressed as an odds ratio $\mathrm{OR}_F = e^{\beta_F}$, representing the multiplicative change in switching odds for a one-standard-deviation increase in $F_k$. To test whether switching depends on the relative timing of the force peak irrespective of direction, we also computed the point-biserial correlation ($r_{\mathrm{pb}}$) between switching events $y_k$ and force peak phase $\phi_{F_k}$ \cite{kornbrot2014point}.

\subsubsection{Model predictions}

Across all dogs, the logistic models showed only weak predictive power for stride-state switching (Table~\ref{tab:amplitude_table}). Areas under the ROC curve ranged from 0.52--0.62, indicating discrimination only slightly above chance ($\mathrm{AUC}=0.5$). Point-biserial correlations between switching and force peak phase were small ($|r_{\mathrm{pb}}|<0.15$), and the fitted circular coefficients $(\beta_{\cos},\beta_{\sin})$ showed no consistent orientation across individuals. Force magnitude contributed minimally, with odds ratios $\mathrm{OR}_F$ close to unity (median $\mathrm{OR}_F = 1.03$), indicating that larger peak forces only marginally increased the probability of a stride-state transition.

Even when the model was fit to the example interval shown in Fig.~\ref{fig:raster}D (45--225\,s), predictive power remained weak ($r_{\mathrm{pb}} = 0.01$, $p_{\mathrm{pb}} = 0.73$, $\mathrm{AUC} = 0.62$, $\mathrm{OR}_F = 1.21$). Consistent with these statistics, no obvious dependence of stride-state switching on phase discrepancy $\Delta\phi_{F_k}$ is evident from Fig.~\ref{fig:raster}D.

\begin{table*}[!t]
\centering
\caption{Summary of logistic model parameters across dogs.}
\begin{tabular}{c r r r r r r r r r}
\hline
ID & $n_{\text{events}}$ & $r_{\text{pb}}$ & $p_{\text{pb}}$ & AUC & Intercept & OR$_F$ & $\beta_{\cos}$ & $\beta_{\sin}$ & $\beta_F$ \\
\hline
1 & 2230 & -0.026 & 0.211 & 0.557 & -1.075 & 1.056 & 0.542 & -0.113 &  0.055 \\
2 & 2225 &  0.076 & 0.000 & 0.586 & -1.030 & 1.037 & 0.317 & -0.698 &  0.036 \\
3 & 2248 & -0.115 & 0.000 & 0.609 & -1.117 & 1.048 & -0.571 &  0.910 &  0.047 \\
4 & 2241 &  0.007 & 0.754 & 0.545 & -0.421 & 0.900 & -0.629 & -0.323 & -0.106 \\
5 & 2213 & -0.147 & 0.000 & 0.619 &  0.028 & 0.780 & -0.516 &  0.417 & -0.249 \\
6 & 2047 &  0.028 & 0.198 & 0.528 & -2.685 & 1.024 & 0.335 & -0.221 &  0.024 \\
7 & 2237 & -0.035 & 0.096 & 0.578 & -0.903 & 0.857 & -0.363 &  0.173 & -0.155 \\
8 & 2329 & -0.028 & 0.176 & 0.522 & -2.983 & 1.257 & -1.018 &  0.212 &  0.229 \\
9 & 2107 & -0.028 & 0.200 & 0.581 &  0.150 & 0.824 & -1.017 & -0.360 & -0.193 \\
10 & 2007 & -0.043 & 0.054 & 0.579 & -1.040 & 1.431 & 0.206 &  0.699 &  0.358 \\
\hline
\end{tabular}
\label{tab:amplitude_table}
\end{table*}

\section{Discussion}

Across thousands of strides, 
each dog occupied a small repertoire of recurrent, acceleration-defined stride states and switched among them frequently during a continuous run. To identify those states without manual gait labels, we combined multivariate dynamic time warping with embedding and density-based clustering. Unlabeled stride classes separated into distinct clusters, confirming that the embedding captured meaningful gait differences rather than noise. Applying this framework to a 10-dog team revealed multiple recurrent stride states during sustained galloping.

These results suggest that sprint sled dog-gallop is not represented by a single invariant acceleration trajectory. Instead, dogs moved  among a  small number of recurrent stride states, often after only a few strides. Dogs frequently transitioned between these states even while maintaining relatively steady speeds and stride periods. Such switching may reflect intrinsic flexibility in limb coordination, allowing animals to maintain stability while running at high speeds.

One plausible hypothesis was that such frequent state switching are driven by local mechanical forcing associated with load pulling. Indeed, mechanical perturbations have been shown to induce gait transitions in several quadrupedal systems \cite{Farley1991trigger,vahedipour2018uncovering}. The specific tugline-force features tested here, however, had limited predictive value. When we examined the relationship between switching events and dynamic features of tugline tension, local mechanical variables around stride boundaries explained little of the observed variation in switching probability. These results indicate that local tugline-force peak features---specifically single-stride peak timing and magnitude---are insufficient predictors of stride-state transitions.

The present data do not identify what drives switching. Mechanical influences not captured here remain plausible---including multi-stride force history, speed fluctuations, and neckline tension between adjacent dogs---and characterizing their role represents a natural direction for future work. We note however, that galloping dogs have been shown to tightly regulate limb timing even on rough terrain \cite{wilshin2020dog}, and stance durations in sprint sled dogs remain highly regular during galloping \cite{ding2025gait}, inconsistent with slip-induced stride-state transitions.

One alternative is that some switches reflect internally generated adjustments in coordination or limb loading rather than immediate responses to local external forcing. In other unloaded galloping quadrupeds,  changes in lead or  asymmetrical coordination have been proposed as a mechanism for balancing muscular workload and mitigating fatigue~\cite{hildebrand1959motions, walter2007ground, wei2015critical}. A similar process may occur here, where flexible stride selection allows dogs to redistribute mechanical effort while sustaining high output over long distances.

Despite the mechanical coupling imposed by the tugline network, stride-state switching also appeared largely individual rather than coordinated across the team. Team-wide switching statistics were consistent with weak and transient coordination, if any. Individual dogs exhibited widely varying dwell times between transitions, with coefficients of variation indicating irregular and burst-like switching rather than periodic alternation. These observations suggest that while sled dogs operate within a mechanically linked system, stride-state selection remains primarily an individual locomotor process.

Several methodological limitations should be noted. First, the unsupervised clustering procedure frequently identified more than four stride states, exceeding the four canonical galloping patterns described in the labeled dataset. One likely explanation is the difference in segmentation conventions. Manually labeled strides were segmented between successive hind-limb touchdown events, whereas the marker-event segmentation used here relied on an arbitrary signal feature closest to the touchdown of the trailing forelimb. This difference may split canonical stride cycles into mixed or transitional segments. Consistent with this interpretation, inspection of transition networks (Fig.~\ref{fig:DTW2}D) shows that several identified states lack strong self-transitions and appear primarily as intermediate states between more stable stride states or gait patterns.

\begin{figure*}[t!]
\centering
\includegraphics[width=0.9\textwidth]{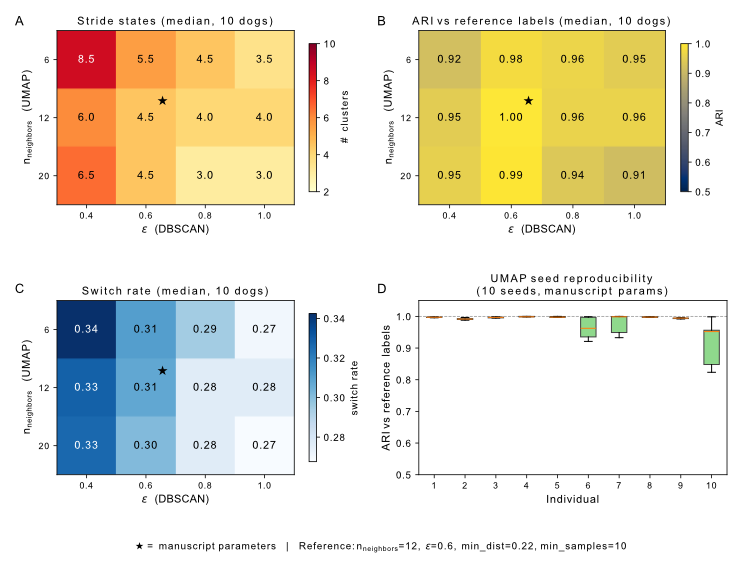}
\caption{\label{fig:robustness} \textbf{Parameter robustness of the UMAP--DBSCAN clustering pipeline.} Heatmaps show median values across 10 dogs as a function of UMAP $n_{\mathrm{neighbors}}$ (rows) and DBSCAN $\epsilon$ (columns), with $\mathrm{min\_dist}$ and $\mathrm{min\_samples}$ held at manuscript values. \textbf{(A)}~Number of identified stride states. \textbf{(B)}~Adjusted Rand Index (ARI) of stride labels relative to the manuscript labeling; values near 1 indicate near-identical stride assignments. \textbf{(C)}~Stride-state switch rate. $\bigstar$~marks the manuscript parameter combination ($n_{\mathrm{neighbors}}=12$, $\epsilon=0.6$). \textbf{(D)}~Seed reproducibility: per-individual ARI between each of 10 random UMAP seeds and the manuscript (seed-42) labeling, at manuscript parameters. Box shows interquartile range; individuals ordered as in Table~\ref{tab:dwell_stats}.}
\end{figure*}

Second, our segmentation assigns a single label per stride at the marker-event timing $t_k$, whereas true transitions may occur at sub-stride resolution. This temporal discretization may blur causal relationships with higher-resolution predictors such as force phase $\phi_{F_k}$. Additionally, UMAP embeddings were computed separately for each dog and are stochastic, preventing direct cross-individual clustering of stride states. Repeating the pipeline under 10 random seeds at the manuscript parameters confirmed that stride-state assignments were highly reproducible (median ARI $= 0.997$), though three individuals showed modestly greater seed sensitivity, likely reflecting less clearly separated stride clusters in those animals (Fig.~\ref{fig:robustness}, IDs 6, 7, and 10). While joint embeddings were explored, individual variation dominated the embedding structure, separating dogs rather than their stride states. Some of this variation may reflect harness fit or sensor placement, which could potentially be reduced using alignment approaches such as Procrustes transformations \cite{gower1975generalized}. However, genuine morphological differences among dogs likely also contribute to the observed variation. Finally, only tugline forces were measured. Neckline forces between adjacent dogs were not recorded and may represent additional channels of mechanical interaction that could influence stride-state selection.

\section{Conclusion}

The unsupervised  pipeline scaled from a single labeled dog   to long field recordings across a 10-dog team and identified recurrent, acceleration-defined stride states. Similar to other unsupervised gait clustering approaches~\cite{Dewhirst_2017}, this framework enables the identification of locomotor modes that are difficult to annotate manually from video or sensor traces alone. In practice, this approach allows extended monitoring of locomotor dynamics across entire sled dog teams, potentially providing new tools for studying performance, training, and health in working animals. More broadly, the approach shows that wearable accelerometry can recover within-gait structure in field settings where continuous side-view video is impractical.

Rather than maintaining a single stereotyped gallop pattern~\cite{Hildebrand1977,Hilderbrand1989Quadrupedal,Hoyt1981gaitenergetic}, sprint sled dogs occupied a small repertoire of stride states and switched among them frequently. These switching events showed little evidence of strong team-wide synchrony, and their timing was only weakly predicted by local tugline-force peak timing and magnitude; broader mechanical influences, such as multi-stride force history or neckline tension, were not assessed here.


Future work that pinpoints sub-stride switch timing, aligns stride spaces across individuals, and incorporates a more complete force network will further clarify the mechanisms underlying these transitions. Controlled comparisons of loaded and unloaded running, ideally alongside physiological measures of effort or fatigue, will be needed to test whether rapid stride-state switching is specific to load pulling or a more general feature of high-speed canine gallop. Applying the same pipeline to datasets that include experimentally controlled acceleration, deceleration, and turning would further clarify how operating conditions influence stride-state selection and switching dynamics, and may expand the repertoire of identifiable stride states beyond those represented here. Direct validation against synchronized video and labeled footfall data during extended field runs would further establish how reliably acceleration-defined stride states correspond to canonical gait patterns across individuals.

Finally, the unexpectedly high prevalence of stride switching raises intriguing questions. Lead changes during galloping have been documented in many quadrupeds \cite{hildebrand1959motions, hackert2008limb}, but appear to occur relatively infrequently. For example, horses galloping on a straight path have been reported to switch transverse lead roughly every $48 \pm 20$ strides \cite{biancardi2012optimal}. By contrast, the stride-state transitions observed in our sled dogs occurred far more frequently, with several individuals exhibiting mean dwell periods of fewer than four strides between transitions. Determining whether this reflects an adaptation to sustained load pulling or a more general feature of high-speed canine locomotion will require future studies of tethered and untethered running dogs.


\section{Animal Subjects and Ethics}
The data analyzed in this study were collected from ten sprint-racing sled dogs from a professional sprint-racing kennel. Dogs were bred and trained for the sport of sprint racing and are representative of the Eurohound type. For the run analyzed here, dog positions within the team were chosen by their owner based on each dog's temperament and aptitude. All procedures were conducted in accordance with institutional and national guidelines for research involving animals. Protocols were reviewed and approved by the Institutional Animal Care and Use Committee at Georgia Institute of Technology (protocol BHAMLA-A100575U-10/06/2025). Written informed consent was obtained from the dogs' owners prior to participation. Dogs were housed and cared for by their owners according to standard husbandry practices. All experimental sessions were integrated into the dogs' normal training routine and were discontinued immediately if any signs of distress were observed.\\


\appendix

\section{Appendix: Parameter Robustness}\label{app:robustness}

To assess sensitivity of the clustering results to the choice of UMAP and DBSCAN parameters, we swept $n_{\mathrm{neighbors}} \in \{6, 12, 20\}$, $\texttt{min\_dist} \in \{0.10, 0.22, 0.40\}$, $\epsilon \in \{0.4, 0.6, 0.8, 1.0\}$, and $\texttt{min\_samples} \in \{4, 10, 20\}$, holding the remaining parameters at manuscript values for each sweep. For each combination, the full pipeline was applied to all ten dogs and three quantities were recorded: number of identified stride states, stride-state switch rate, and the Adjusted Rand Index (ARI) of the resulting stride labels relative to the manuscript labeling. ARI compares all pairwise stride relationships between two labelings, adjusted for chance, such that ARI~$= 1$ indicates identical assignments and ARI~$\approx 0$ indicates agreement no better than chance.

Across the swept parameter space, the median number of stride states was stable in the neighborhood of the manuscript parameters (median 5, range 3--9; Fig.~\ref{fig:robustness}A). The cluster count was most sensitive to $\epsilon$, as expected: very small values cause DBSCAN to over-fragment the embedding, while large values merge distinct clusters. The manuscript value ($\epsilon = 0.6$) sits at the transition between these regimes. Stride-state switch rate was notably robust, varying between 0.27 and 0.34 across all tested parameter combinations, indicating that the central finding of frequent switching is insensitive to parameter choice (Fig.~\ref{fig:robustness}C). Stride-level cluster assignments were also highly consistent with the manuscript labeling. The median ARI across non-reference parameter combinations was 0.95, with a lower quartile of 0.91 (Fig.~\ref{fig:robustness}B).

UMAP's stochastic initialization was assessed separately by repeating the pipeline under 10 random seeds at the manuscript parameters. Stride-state assignments were highly reproducible across seeds (median ARI~$= 0.997$; minimum $= 0.698$). Three individuals showed modestly greater seed sensitivity (Fig.~\ref{fig:robustness}D), likely reflecting embedding geometries in which clusters sit in closer proximity, or less stereotyped individual acceleration profiles, making assignments near cluster boundaries sensitive to small changes in initialization.

\section{Competing Interests}
No competing interest is declared.

\section{Author Contributions}

Author contributions: Conceptualization:
B.S. Experiments and related data analysis: B.S.
Model and related analysis: B.S. Writing---original
draft: B.S. Writing---review and editing: B.S., S.B. Visualization: B.S. Supervision, funding acquisition: S.B.

\section{Acknowledgments}
We gratefully acknowledge the generous support of the sled dog mushers who volunteered their kennel and dogs for data collection. Their willingness to share their time, expertise, and hospitality made this work possible. We thank M. Bull, H. Huson, and Y. Chang for helpful discussions that contributed to this work.  S.B. acknowledges funding from NSF CAREER IOS-1941933 and Schmidt Sciences, LLC. 

\section{Data and Materials Availability}
Custom sensor design files are available at \href{https://github.com/bhamla-lab/DoggyLogger}{\texttt{https://github.com\\/bhamla-lab/DoggyLogger}}.
Code used to generate all manuscript figures, along with processed data is available at \href{https://github.com/bhamla-lab/sled-dog-switching}{\texttt{https://github\\.com/bhamla-lab/sled-dog-switching}}.


\end{document}